\pdfoutput=1
\documentclass[a4paper,fleqn,usenatbib]{mnras}

\usepackage{mathptmx}

\usepackage[T1]{fontenc}
\usepackage{ae,aecompl}

\usepackage{graphicx}    
\usepackage{amsmath}    
\usepackage{amssymb}    

\newcommand{\Mpc}{\,\mathrm{Mpc}}    
\newcommand{\kpc}{\,\mathrm{kpc}}    
\newcommand{\M}{\,\mathrm{M}}    

\title[Massive subhaloes in clusters]{Resolution of the apparent
  discrepancy between the number of massive subhaloes in Abell~2744
  and $\Lambda$CDM }

\author[Tian-Xiang Mao et al]{
\parbox[t]{\textwidth}{Tian-Xiang Mao$^{1,2}$\thanks{Email: maotianxiang@bao.ac.cn }, Jie Wang$^{1}$, Carlos S. Frenk$^{3}$, Liang Gao$^{1}$, Ran Li$^{1,4}$,\\ Qiao Wang$^{1}$, Xiaoyue Cao$^{1,2}$ and Ming Li$^{1}$}\\ \\
\parbox[t]{\textwidth}{
$^{1}$Key laboratory for Computational Astrophysics, National Astronomical Observatories, Chinese Academy of Sciences, Beijing, 100012, China\\
$^{2}$University of Chinese Academy of Sciences, Beijing 100049,China\\
$^{3}$Institute for computational cosmology, Department of Physics, University of Durham, South Road, Durham, DH1  3LE, UK\\
$^{4}$College of Astronomy and Space Sciences, University of Chinese Academy of Sciences, Beijing 100049, China
} }

\date{Accepted 2018 April 17. Received 2018 April 16; in original form 2017 August 04}

\pubyear{2015}

\begin{document}
\label{firstpage}
\pagerange{\pageref{firstpage}--\pageref{lastpage}}
\maketitle
 
\begin{abstract}
  Schwinn et al. (2017) have recently compared the abundance and 
  distribution of massive substructures identified in a gravitational lensing 
  analysis of Abell 2744 by Jauzac et al. (2016) and N-body simulation and 
  found no cluster in $\Lambda$CDM simulation that is similar to Abell~2744.
  Schwinn et al. (2017) identified the measured projected aperture masses with 
  the actual masses associated with subhaloes in the MXXL N-body
  simulation. We have used the high resolution Phoenix cluster
  simulations to show that such an identification is incorrect: the
  aperture mass is dominated by mass in the body of the cluster that
  happens to be projected along the line-of-sight to the subhalo. This
  enhancement varies from factors of a few to factors of more than
  100, particularly for subhaloes projected near the centre of the
  cluster. We calculate aperture masses for subhaloes in our
  simulation and compare them to the measurements for Abell~2744. We
  find that the data for Abell~2744 are in excellent agreement with
  the matched predictions from $\Lambda$CDM. We provide further
  predictions for aperture mass functions of subhaloes in idealized
  surveys with varying mass detection thresholds.
\end{abstract}

\begin{keywords}
   cold dark matter -- gravitational lens -- cluster - substructure 
\end{keywords}



\section{Introduction}

The existence of a very large number of dark matter haloes and subhaloes is a
fundamental prediction of the $\Lambda$CDM cosmology. The halo and
subhalo mass functions can be accurately calculated from N-body
simulations
\citep[e.g.][]{Frenk_1988,Jenkins_2001,Gao_2004,Gao_2011,Gao_2012,Springel_2008}. These
functions are characteristic of CDM and can differ in models with
different types of dark matter such as warm or self-interacting dark
matter \citep{hellwing_2016,bose_2016,Vogelsberger_2012}.

The abundance of haloes and subhaloes, their mass, and their spatial
distribution can, in principle, be measured from their weak gravitational
lensing effects \citep[e.g.][]{Yang_2006,Natarajan_2007, Natarajan_2009, 
Limousin_2007, Okabe_2014, Li_2013, Li_2015,Li_2016}. 
Strong lensing may be used to
measure the small mass end of the mass function \citep{Vegetti_2008},
and this provides a promising test to differentiate between, for
example, cold and warm dark matter \citep[e.g.][]{Li_2016,Li_2017}. The
combination of strong and weak lensing may be used to measure the mass
function on larger mass scales. 
Using this approach \cite{Jauzac_2015,Jauzac_2016} have reconstructed the total projected  mass distribution of Abell~2744, one of the most massive galaxy clusters known (which lies at $z = 0.308$).
Their technique is particularly
sensitive to density variations in the outer parts of the cluster and
thus is ideal for identifying subhaloes in these regions and estimating
their mass quite accurately.

\cite{Jauzac_2016} identified seven massive subhaloes (or eight if the
main core is included) within a radius of $1 \Mpc$ around the centre
of Abell 2744; they estimated their enclosed mass within an aperture
of $R=150~\kpc$ to be greater than $5\times
10^{13}~\mathrm{M}_{\odot}$ in all cases. 
\cite{Schwinn_2017} claimed that this result is inconsistent with 
the abundance and distribution of cluster subhaloes in the Millennium XXL 
simulation \citep[MXXL;][]{Angulo_2012} and that a more careful comparison 
should be performed.
They considered subhaloes previously
identified in a sample of clusters in the simulation and, assuming
that they have NFW density profiles \citep{NFW}, they estimated their
mass by integrating the density profile within a cylindrical volume of
radius $R = 150~\mathrm{kpc}$ and length $l=30~\mathrm{Mpc}$, finding
a maximum of three subhaloes with mass $\mathrm{M}(R<150\
\mathrm{kpc})>5\times 10^{13}\ \mathrm{M}_{\odot}$ located within
$1~\mathrm{Mpc}$ of the centre. They attempted to account for possible
effects, such as projection along the line-of-sight or changes in the
assumed subhalo density profiles induced by baryons but found these to
be unimportant and concluded that the number of observed massive
subhaloes in Abell~2744 is in conflict with the predictions of
$\Lambda$CDM.

To try and explain the discrepancy, \cite{Lee_2017} hypothesized that
Abell~2744 may be embedded within a filamentary supercluster aligned
with the line-of-sight. \cite{Natarajan_2017} compared the subhalo
mass function of galaxy members in Abell~2744 with the results for
clusters in a hydrodynamical simulation \citep{Vogelsberger_2014}, finding no
discrepancy between observations and the simulation.  The discrepancy,
however, does not appear to be exclusive to
Abell~2744. \cite{Chiriv_2017} found a similar mismatch with N-body
simulations in the MACS J0416.1-2403 cluster and, like
\cite{Schwinn_2017}, found that projection effects cannot account for
the discrepancy.  

In this paper we show that the discrepancy between Abell~2744 
and the $\Lambda$CDM simulation  reported by \cite{Schwinn_2017} is simply 
due to an inconsistency in their comparison with the MXXL simulation, 
specifically an inconsistency between the masses they infer for the 
subhaloes and the masses assigned to subhaloes in the simulation.
We mimic the procedure of deriving an aperture mass that was applied to
the lensing data in the high-resolution Phoenix \citep{Gao_2012} and
Indra (Falck et al., in prep.) N-body simulations and find that the
discrepancy with Abell~2744 is removed.

In Section~\ref{sec:data} we describe the Phoenix simulations; the
comparison with Abell~2744 is presented in
Section~\ref{sec:result}. Finally, we discuss our results and draw
conclusions in Section~\ref{sec:con_dis}.

\section{Simulations}
\label{sec:data}

The N-body simulations used in this study are the Indra suite of large
cosmological simulations (Falck et al., in prep.) and the Phoenix set
of very high resolution simulations of individual rich clusters
\citep{Gao_2012}. 

Indra consists of 512 N-body simulations, each with $1024^3$ dark
matter particles in a periodic cube 1~$h^{-1}\mathrm{Gpc}$ on a
side. The cosmological parameters are taken to be: $\Omega_{m}=0.272$,
$\Omega_{\Lambda}=0.728$, $\Omega_{b}=0.045$, $h=0.704$,
$\sigma_{8}=0.81$, and $n_{s}=0.967$.  Indra includes a very large
volume and thus produces a large sample of clusters like Abell~2744,
but the resolution, $m_p=7.03 \times 10^{10}~
h^{-1}\mathrm{M}_{\odot}$, is too low to resolve subhaloes like those
in Abell~2744.  We use these simulations to compare aperture against
total masses for clusters analogous to Abell~2744.

Phoenix consists of very high resolution resimulations of nine clusters
and their surroundings selected from the Millennium simulation
\citep{Springel_2005}. The Millennium simulation assumes cosmological
parameters consistent with the first year WMAP data,
$\Omega_{m}=0.25$, $\Omega_{\Lambda}=0.75$, $\sigma_{8}=0.9$,
$n_{s}=1$ and $h=0.7$. These values deviate from the latest Planck
results but this small offset is of no consequence for the topic of
this study. The most massive of the nine Phoenix clusters, `Ph-I', has a
virial mass of $M_{200}=2.427\times 10^{15}
~h^{-1}\mathrm{M}_{\odot}$, close to that of Abell 2744 (see
Section~\ref{sec:result}), and we choose this halo at $z=0.32$ for detailed comparison
with Abell~2744. We identify subhaloes in Ph-I using the SUBFIND
algorithm of \cite{Springel_2001}. To test numerical convergence, the
Phoenix clusters were re-simulated at various resolutions. In this
study we have analyzed the `level-4' resolution for which the particle
mass is $m_p=4.559 \times 10^{8}~h^{-1}\mathrm{M}_{\odot}$; at this
resolution all massive subhaloes are well resolved.


\section{Results}
\label{sec:result} 
\subsection{The aperture mass of Abell 2744}

\begin{figure}
    \begin{center}
        \includegraphics[width=1.0\columnwidth]{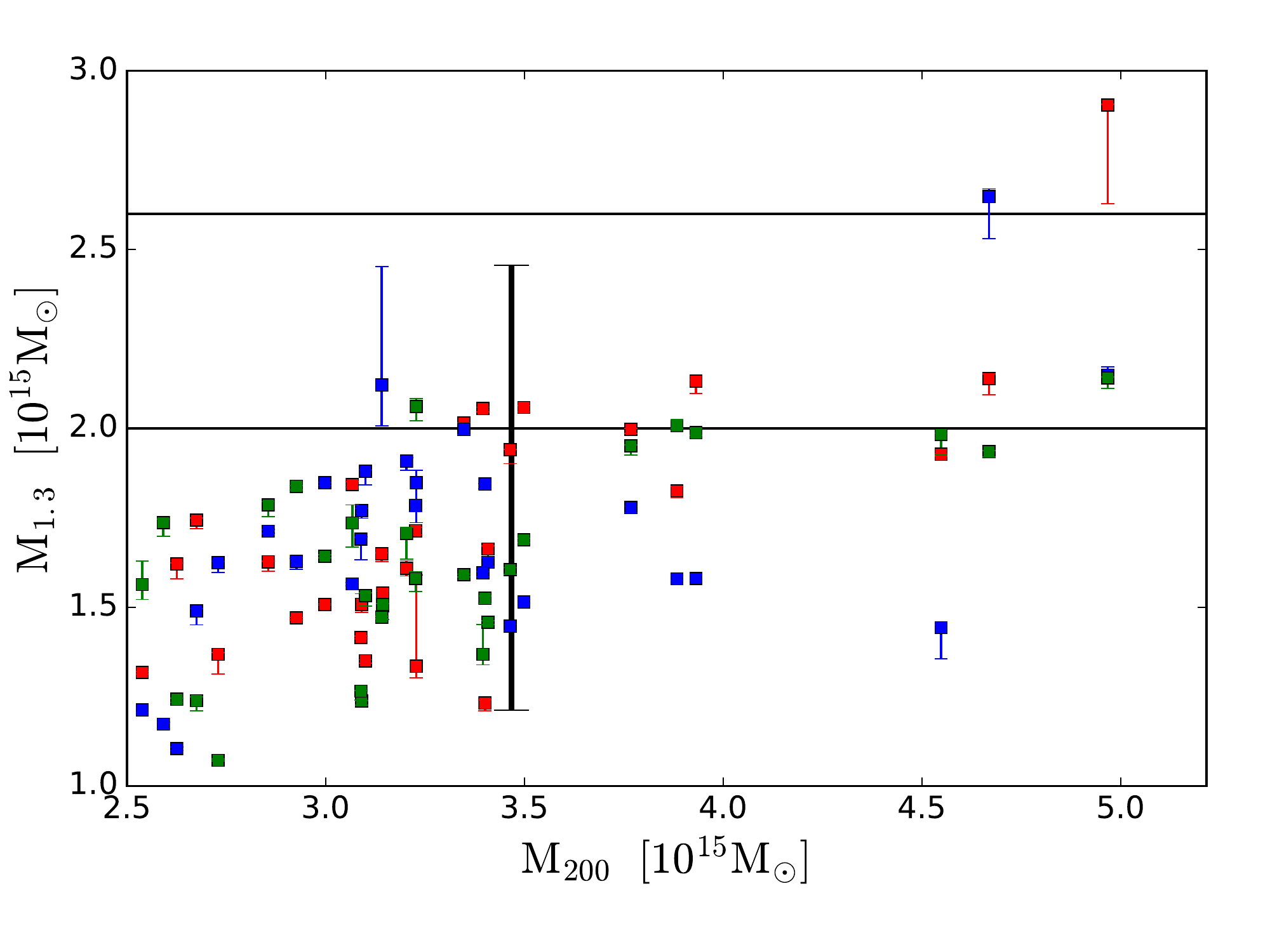}
    \end{center}
    \caption{The relation between the projected aperture mass of a
      halo, $M_{1.3}$, calculated within a projected cylinder of
      radius $r= 1.3~\Mpc$ and depth $30~\Mpc$, and its
      virial mass, $M_{200}$.  Each point shows a projection of a
      cluster in the Indra simulations, with red, green and blue
      indicating three orthogonal projects. The upper and lower error
      bars show $60$ and $10~{\rm Mpc}$ projection depths,
      respectively.  The black error bar shows the entire range of
      $M_{1.3}$ in 200 random projections of Ph-I-4. The $3\sigma$
      mass range in $M_{1.3}$ for Abell 2744 is marked by the two
      solid horizontal lines. }
    \label{fig:ProjM}
\end{figure}

Since the subhalo mass function depends on the host halo mass
\citep[e.g.][]{Gao_2011}, it is important to select simulated clusters
of mass similar to that of Abell~2744. There are different ways to
define the mass of a halo in a simulation, but in lensing analyses 
the mass of the lens is usually estimated as the projected mass
within a certain circular aperture. For Abell~2744, the
aperture has radius $R=1.3~\Mpc$.

We select all haloes of mass, $M_{200}>2.5\times 10^{15}
~\mathrm{M}_{\odot}$ from 128 realizations of the Indra simulation
suite (a volume equivalent to a cube of side $128
h^{-3}~\mathrm{Gpc}^{3}$) at $z=0.32$. For each halo, we compute an
`aperture mass', analogous to that Abell~2744, by projecting the
particle distribution along the $x$-, $y$-, and $z$-axis of the
simulation, keeping all dark matter particles within $1.3\Mpc$ of the
centre of each cluster. The depth of projection was chosen to be $10$,
$30$ and $60~\Mpc$. We refer to all these masses collectively as
$M_{1.3}$. In Fig.~\ref{fig:ProjM} we plot
$M_{200}$\footnote{$M_{200}$ is defined as the mass contained with the
  radius, $r_{200}$, at which the mean interior density is equal to
  200 times the critical density.} against aperture mass, $M_{1.3}$,
for the clusters in our sample.  Different colours represent the three
different projections; the error bars indicate the scatter in aperture
mass due to different projection depths. The projection depth has a
negligible effect on the aperture mass except in a few cases where there is
contamination by a massive structure along the line-of-sight.  The
aperture mass is thus essentially insensitive to the assumed
projection depth.

The observed $3 \sigma$ range of $M_{1.3}$ for Abell 2744 is marked by
the two solid horizon lines in Fig.~\ref{fig:ProjM}. To fall in the
allowed region of $\M_{1.3}$ for Abell~2744, a cluster should have
$M_{200}>3 \times 10^{15} \M_{\odot} $. Only one of the nine Phoenix
clusters, Ph-I, has such a large mass; the range of values of
$M_{1.3}$ for that cluster from 200 random projections is indicated
with a black error bar. Only ten percent of these projections fall
within the $3\sigma$ allowed region of $M_{1.3}$ for Abell~
2744. However, since the amplitude of the subhalo mass function scales
approximately linearly with halo mass \citep{Wang_2012}, our
conclusions from comparing this simulated cluster (the only one
available with the required resolution) with Abelll~2744 are
conservative.

\begin{figure}
    \begin{center}
        \includegraphics[width=0.9\columnwidth]{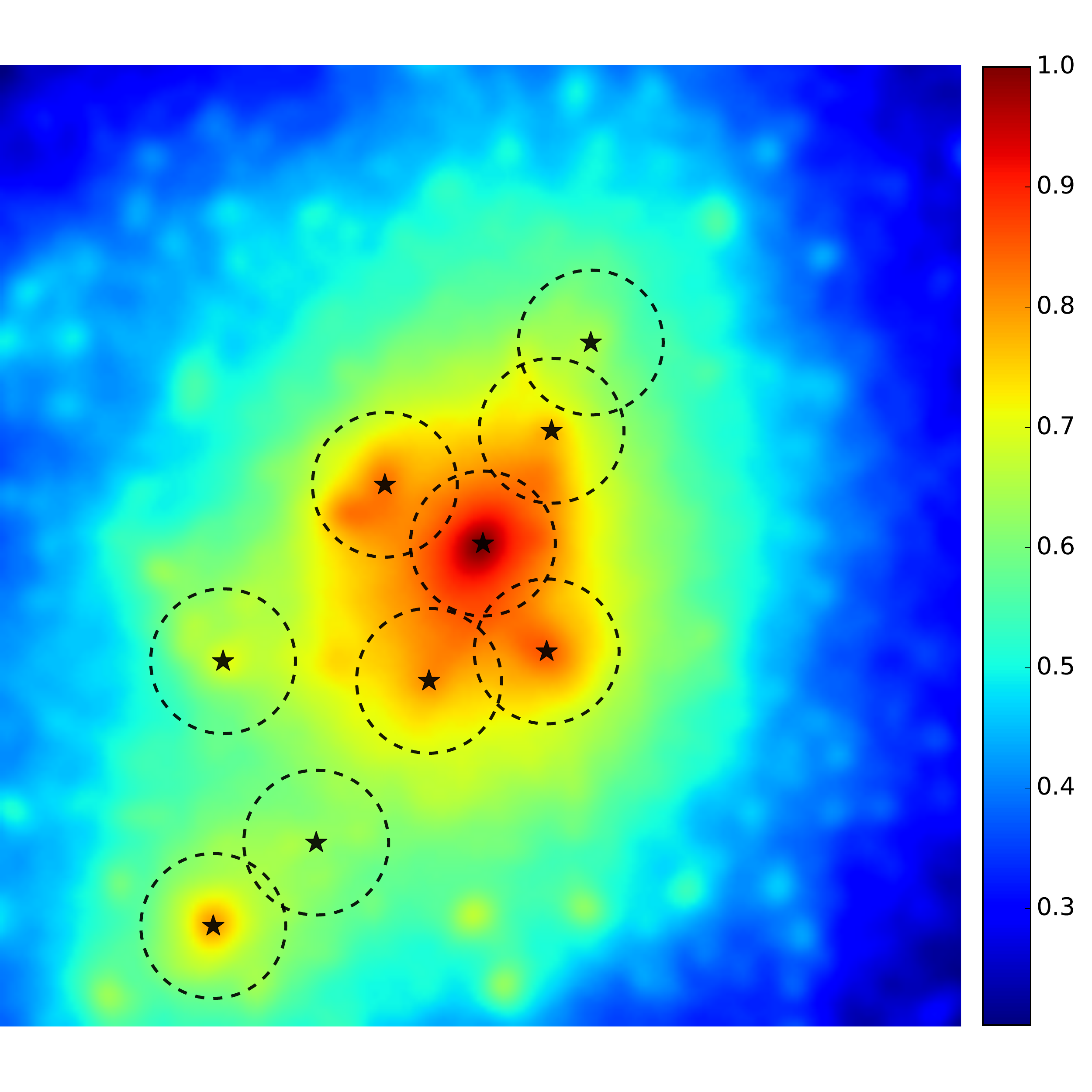}
    \end{center}
    \caption{A $2 \Mpc \times 2 \Mpc$ projected mass map of Ph-I-4
      viewed along a random direction. The black stars, surrounded by
      circles of radius $R =150 ~{\mathrm{kpc} }$, mark substructures
      with aperture mass, $\mathrm{M}_{150} > 5\times 10^{13}
      ~\mathrm{M}_{\odot}$.  The mass map is in a logarithmic scale.
    }
    \label{fig:massmap}
\end{figure}

\subsection{Projected massive subhaloes in clusters}

We project the particle distribution of Ph-I-4 along 200 random
projections, each of depth $30~\Mpc$. The resulting mass maps are not
sensitive to the projection depth as long as it is greater than the
diameter of the cluster.  In $24$ of these projections the Ph-I
cluster has aperture mass, $M_{1.3}$, within the 3$\sigma$ allowed
range for Abell~2744.  Hereafter, we refer to these as our Abell~2744
analogues, which use to compare the simulation with the observational
data.

In lensing observations, subhalo candidates are identified in the
reconstructed mass map. In this paper, we will assume that all
subhaloes of mass larger than a threshold, $M_{\rm th}$, are detected
in the lensing analysis. For each massive subhalo in the simulation we
calculate an aperture mass, $M_{150}$, analogous to the aperture
masses measured in observational analyses \citep[e.g][]{Jauzac_2016}
by measuring the mass that falls within a projected radius, $R =
150~\mathrm{kpc}$.  Since close subhalo pairs cannot be distinguished
in lensing observations, we merge the density peaks of subhalo pairs
of separation less than $200~\kpc$, which is approximately the
shortest pair separation among the massive subhaloes in Abell~2744.

In Fig.~\ref{fig:massmap}, we show the mass map of the particular
projection that has the most abundant substructures among the 200
projections of Ph-I-4. Setting a subhalo mass detection limit of
$M_{\rm th}=4.6\times 10^{11} ~\mathrm{M_{\odot}}$, we find nine
subhaloes (shown as black stars) whose aperture masses are comparable
to the aperture masses of the subhaloes in Abell 2744. It is clear
that some of these apparently massive subhaloes in projections are
actually associated with rather puny subhaloes such as one of the pair
in the bottom left of the image or the one slightly above that 
on the right.

\begin{figure}
    \begin{center}
        \includegraphics[width=1.0\columnwidth]{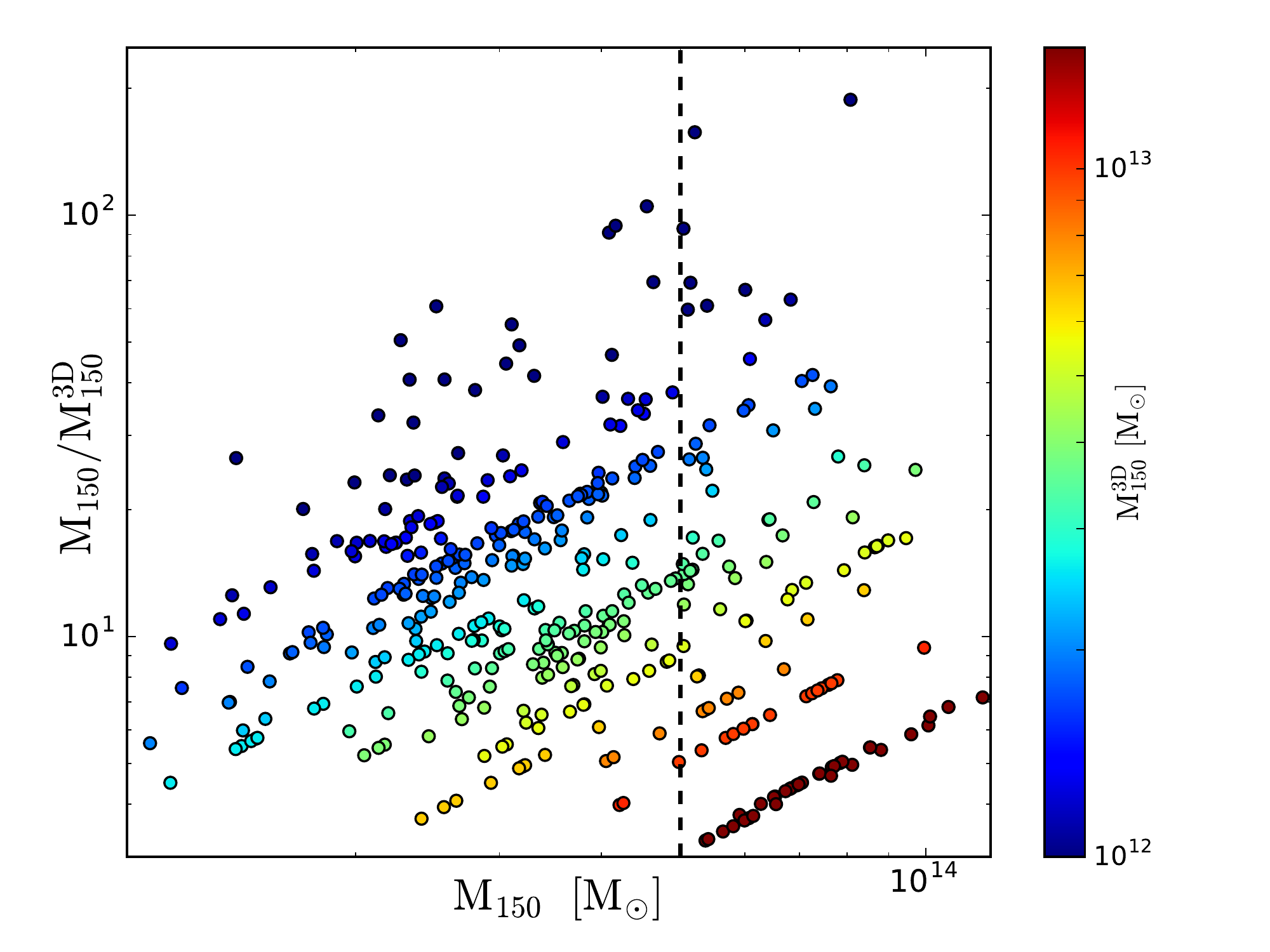}
    \end{center}
    \caption{The relation between the aperture mass of subhaloes,
      $M_{150}$, and their mass `boost' factor,
      $M_{150}/M^{3\mathrm{D}}_{150}$, in our 24 Abell~2744
      analogues. Here, $M^{3\mathrm{D}}_{150}$ is the mass enclosed
      within a sphere of radius $150~\kpc$ of the subhalo centre. The
      points are color coded according to the value of
      $M^{3\mathrm{D}}_{150}$. A dashed vertical line marks the limit
      $M_{150}=5\times 10^{13} ~\mathrm{M_{\odot}}$.  }
    \label{fig:enhancement}
\end{figure}

In Fig.~\ref{fig:enhancement} we show the relation between the
aperture mass, $M_{\rm 150}$, measured from different directions, and
the mass `boost factor', $M_{\rm 150} / M_{\rm 150}^{\rm 3D}$, where
$M_{\rm 150}^{\rm 3D}$ is the true mass of the subhalo contained
within a sphere of radius $150~\kpc$ around the subhalo centre
identified with SUBFIND.  Clearly, the aperture mass, of a subhalo can
be very different from its real mass, $M_{\rm 150}^{\rm 3D}$. This is
because aperture masses can be greatly boosted by mass in the body of
the halo which happens to fall within the projection. This can
increase the projected mass by factors varying from a few to about
100. Thus, even intrinsically small subhaloes can appear to be very
massive as judged by their aperture mass, particularly if they happen
to be projected close to the host halo centre.

\begin{figure}
    \begin{center}
        \includegraphics[width=1.0\columnwidth]{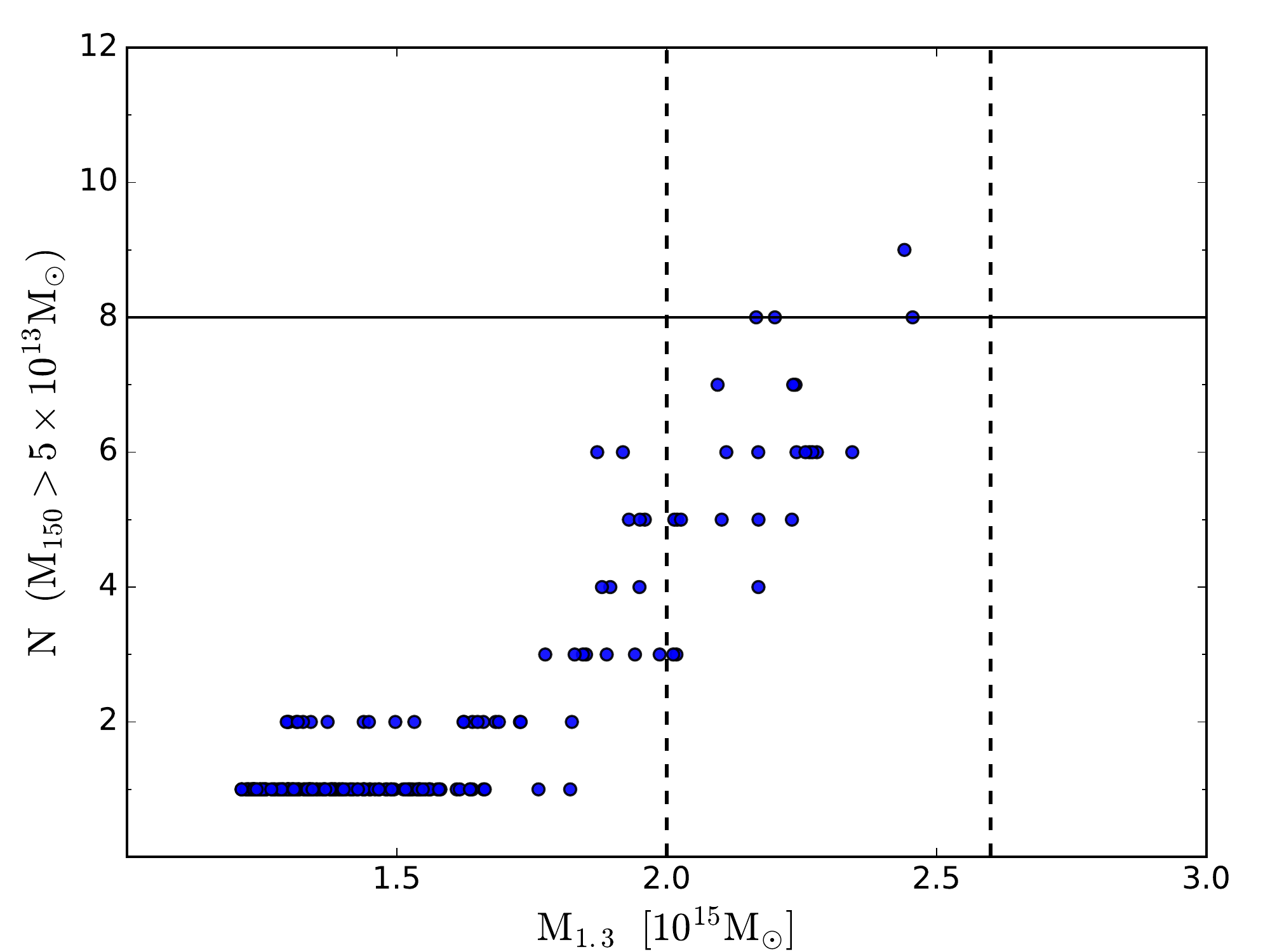}
    \end{center}
    \caption{The number of massive substructures, $M_{150}> 5\times
      10^{13} ~\mathrm{M_{\odot}}$, as a function of the aperture mass
      for all 200 projections of Ph-I-4. Each point corresponds to one
      projection. The $3\sigma$ allowed range of $M_{1.3}$ for
      Abell~2744 is shown by the dashed vertical lines. The horizontal
      line corresponds to the eight subhaloes with $M_{150}> 5\times
      10^{13} ~\mathrm{M_{\odot}}$ found in Abell~2744.  }
    \label{fig:N_M}
\end{figure}

In Fig.~\ref{fig:N_M} we show the number of projected subhaloes of
aperture mass, $M_{\rm 150}>5\times 10^{13}~\mathrm{M_{\odot}}$, in
our 200 projections of Ph-I-4, assuming a projected mass detection
limit of $M_{\rm th}=4.6\times 10^{11} \M_{\odot}$.  Clearly, the
number correlates strongly with the projected aperture mass. This is
expected because, as we have just seen, the subhalo aperture mass is
dominated by mass in the body of the cluster that is projected along
the line-of-sight. Among 24 projections, four have at least eight
subhaloes with $M_{150}>5\times 10^{13}~\mathrm{M_{\odot}}$, as
indicated by the horizontal line. We conclude that when the
simulations and the data are analyzed in a consistent way, the
detection of eight massive subhaloes in Abell~2744 is perfectly
consistent with the predictions of $\Lambda$CDM.


\begin{figure}
    \begin{center}
        \includegraphics[width=1.0\columnwidth]{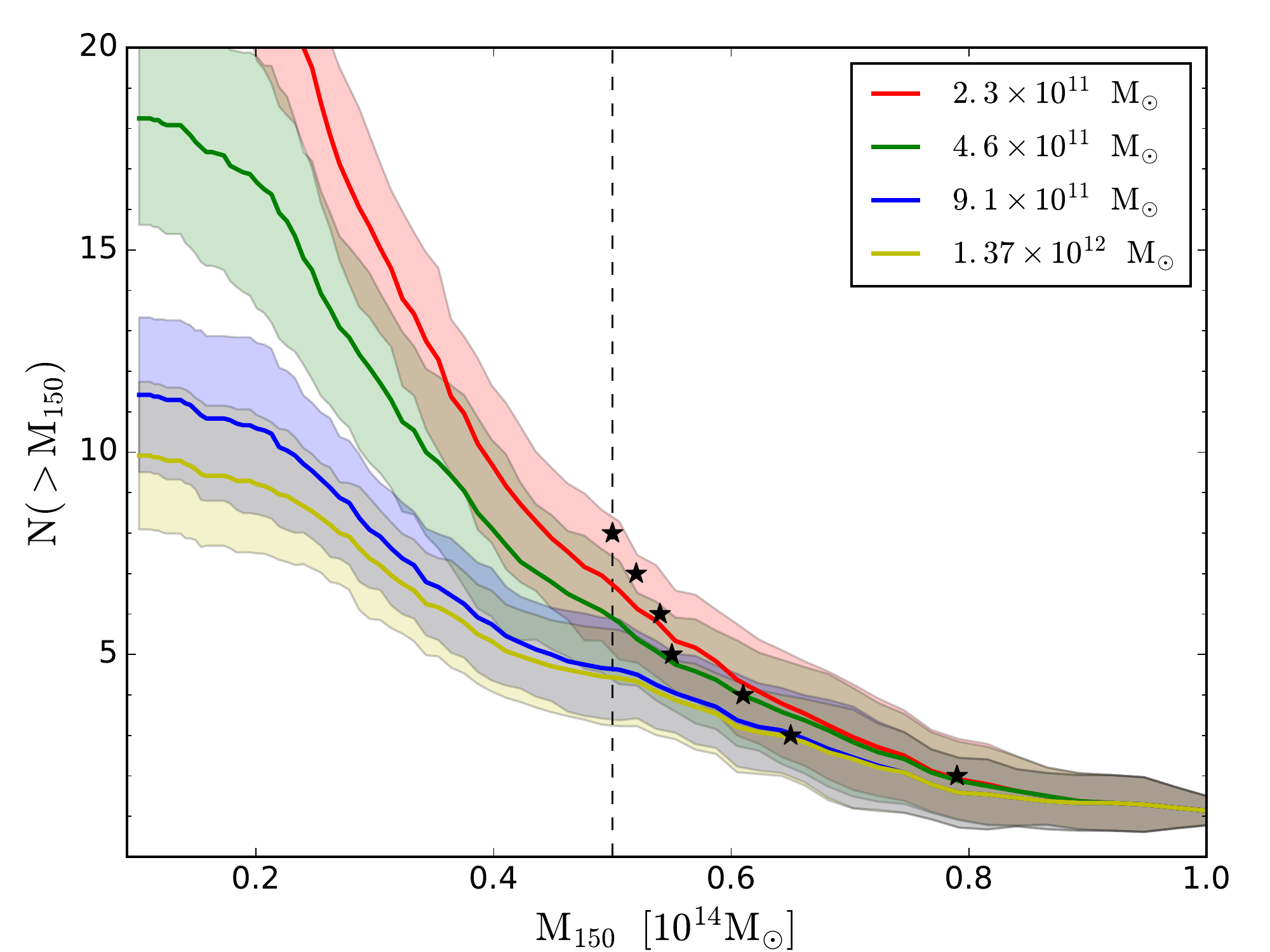}
    \end{center}
    \caption{The predicted aperture mass function of subhaloes.  The
      solid lines show the mean values, and the shadow areas the $1\sigma$
      range, obtained from our 24 analogues of Abell~2744. Different colours
      correspond to threshold values of 2.3, 4.6, 9.1 and 13.7 $\times
      10^{11} ~\mathrm{M_{\odot}}$, respectively, as shown in the
      legend. The black star symbols show the cumulative mass function
      of the 8 massive substructures in Abell~2744.  }
    \label{fig:mf}
\end{figure}

Lowering the detection mass threshold, $M_{\rm th}$, rapidly increases
the number of massive projected subhaloes. In Fig.~\ref{fig:mf} we
show the predicted aperture mass functions of subhaloes in our 24
Abell~2744 analogues for different values of the threshold.  Red,
green, blue and yellow lines correspond to threshold values, $M_{\rm
  th}$ of 2.3, 4.6, 9.1 and 13.7 $\times 10^{11} {\rm M}_{\odot}$,
respectively. For the value, $M_{\rm th}=2.3\times
10^{11}~\mathrm{M_{\odot}}$, that we have assumed for the lensing
analysis of Abell~2744 carried out by \cite{Jauzac_2016}, the measured
aperture mass function (shown as star symbols in the figure) agrees
remarkably well with the simulations.

\section{Discussion and conclusions}
\label{sec:con_dis}

We have made use of cosmological N-body simulations to test whether the 
identification of eight massive subhaloes by \cite{Jauzac_2016} in a 
gravitational lensing mass reconstruction of Abell~2744 is in conflict
with predictions from the $\Lambda$CDM cosmological paradigm.
Gravitational lensing is
sensitive to projected mass; projected masses associated with subhalos
are normally measured within a specified aperture.  Firstly, using a
large-volume, low-resolution suite of simulations we established that
the projected aperture mass of Abell~2744 itself corresponds to a minimum
true mass of about $3 \times 10^{15} \M_{\odot}$. One of the clusters
from the much higher resolution Phoenix cluster N-body re-simulation
project satisfies this mass constraint; we used it to construct a
sample of 24 analogues of Abell~2744 by viewing it from different
directions.

Projected masses for subhaloes in Abell~2744 are measured within
$150~\kpc$ apertures. We calculated equivalent masses for the
subhaloes in the simulation by integrating the mass within a cylinder
of radius $150~\kpc$ along the line-of-sight to each subhalo. Our main
finding is that the measured aperture mass of a subhalo is dominated
by mass in the body of the host halo that happens to be projected onto
the aperture. This can lead to measured aperture masses that are over
100 times larger than the actual mass associated with the subhalo.

Although our procedure captures the main effect of measuring aperture
masses, our comparison with the observations of Abell~2744 is only
approximate. In practice, the detectability of subhaloes in
gravitational lensing analyses depends not only on their mass, but
also on the mass reconstruction method. For example,
\cite{Merten_2011} also performed a strong plus weak lensing analysis
of Abell~2744, using imaging data from HST and Subaru; they were only
able to find 4 of the \cite{Jauzac_2016} subhaloes. A more realistic
comparison with the results of \cite{Jauzac_2016} would require a
ray-tracing calculation allowing for limitations and complications of
the observational data such as resolution, completeness, etc. Such a
calculation is beyond the scope of this paper. 
However, our main result-- that projected aperture masses of subhalos in
observed rich clusters can be significant overestimates of the true masses of the
subhaloes-- is general and sufficient to conclude that the number of
substructures detected in Abell~2744 by \cite{Jauzac_2016} does not
pose a crisis for $\Lambda$CDM. We have presented simple predictions
for the aperture mass function of subhaloes in rich clusters. More
detailed theoretical calculations of the kind we have sketched above, 
tailored to specific lensing surveys, could provide a useful test of
$\Lambda$CDM.

\section*{Acknowledgements}

We thank Dandan Xu, Guoliang Li and Simon White for useful discussions. 
We acknowledges the 973 program grant 2015CB857005, 2017YFB0203300,
and NSFC grant No. 11373029, 11390372, 11851301.
We acknowledge support from 
NSFC grants (nos. 11573030, 11133003, 11425312 and 1303033).
RL acknowledges NSFC grant (Nos. 11511130054, 11333001), 
support from the Youth Innovation Promotion Association of CAS 
and Nebula Talent Program of NAOC and Newton Mobility award.
This work was supported by the Science and
Technology facilities Council ST/L00075X/1. It used the DiRAC
Data Centric system at Durham University, operated by the Institute
for Computational Cosmology on behalf of the STFC DiRAC HPC Facility
(www.dirac.ac.uk). This equipment was funded by BIS National
E-infrastructure capital grant ST/K00042X/1, STFC capital grants
ST/H008519/1 and ST/K00087X/1, STFC DiRAC Operations grant
ST/K003267/1 and Durham University. DiRAC is part of the National
E-Infrastructure. 




\bibliographystyle{mnras}
\bibliography{abell} 



%
%


\bsp    
\label{lastpage}
\end{document}